\documentclass[apj]{emulateapj}

\newcommand{\mc}{\multicolumn}
\newcommand{\gsim}{\gtrsim}

\newcommand{\kms}{\ensuremath{\mbox{km}\,\mbox{s}^{-1}}}
\newcommand{\masyr}{\ensuremath{\mbox{mas}\,\mbox{yr}^{-1}}}

\newcommand{\hst}{\textit{HST}}

\defcitealias{nr06}{NR06}
\defcitealias{cm99}{CM99}
\defcitealias{wm77}{WM77}

\begin{document}

\title{A Precise Proper Motion for the Crab Pulsar, and the 
Difficulty of Testing Spin-Kick Alignment for Young Neutron Stars}

\author{D.~L.~Kaplan\altaffilmark{1}, S.~Chatterjee\altaffilmark{2},
  B.~M.~Gaensler\altaffilmark{2,3}, and J.~Anderson\altaffilmark{4}}

\altaffiltext{1}{Pappalardo Fellow and Hubble Fellow; Kavli Institute for Astrophysics and Space
  Research and Department of Physics, Massachusetts Institute of
  Technology, Cambridge, MA 02139; dlk@space.mit.edu.} 
\altaffiltext{2}{School of Physics, The University of Sydney, NSW
  2006, Australia; schatterjee, bgaensler@usyd.edu.au}
\altaffiltext{3}{Harvard-Smithsonian Center for Astrophysics, Cambridge, MA 02138}
\altaffiltext{4}{Department of Physics and Astronomy, Rice
  University, Houston, TX 77251; jay@eeyore.rice.edu} 

\slugcomment{Accepted for Publication in ApJ}

\begin{abstract}
We present a detailed analysis of archival \textit{Hubble Space
Telescope} data that we use to measure the proper motion of the Crab
pulsar, with the primary goal of comparing the direction of its proper
motion with the projected axis of its pulsar wind nebula (the
projected spin axis of the pulsar).  Combining data from 47
observations spanning $>10\,$yr with two different instruments, and
using the best available measurement techniques and latest distortion
models, we are able to demonstrate that our measurement is robust and
has an uncertainty of only $\pm0.4\,\masyr$ on each component of the
proper motion.  However, we then consider the various uncertainties
that arise from the need to correct the proper motion that we measure
to the local standard of rest at the position of the pulsar and find
$\mu_\alpha = -11.8 \pm 0.4 \pm 0.5\,\masyr$ and $\mu_\delta =
\mbox{+4.4} \pm 0.4 \pm 0.5\,\masyr$ relative to the pulsar's standard
of rest, where the two uncertainties are from the measurement and the
reference frame, respectively.  If we then wish to compare this proper
motion to the symmetry axis of the pulsar wind nebula, we must
consider the unknown velocity of the pulsar's progenitor (assumed to
be $\sim 10\,\kms$), and hence add an additional uncertainty of
$\pm2\,\masyr$ to each component of the proper motion, although this
could be a factor of 10 larger if the pulsar's progenitor had an
anomalously high velocity ($>100\,\kms$).  This implies a projected
misalignment with the nebular axis of $14\degr\pm2\degr\pm9\degr$,
consistent with a broad range of values including perfect alignment.
We use our proper motion to derive
an independent estimate for the site of the supernova explosion with
an accuracy that is 2--3 times better than previous estimates.  We conclude
that the precision of individual measurements which compare the
direction of motion of a neutron star to a fixed axis will often be
limited by fundamental uncertainties regarding reference frames and
progenitor properties. The question of spin-kick (mis)alignment, and
its implications for asymmetries and other processes during supernova
core-collapse, is best approached by considering a statistical
ensemble of such measurements, rather than detailed studies of
individual sources.
\end{abstract}


\section{Introduction}
The high velocity nature of the neutron star 
population has been apparent almost as long as the existence of
neutron stars has been recognized \citep{go70}. Recent
statistical studies of the radio pulsar velocity distribution
\citep[e.g.,][]{acc02,bfg+03,hllk05,fgk06} yield mean
three-dimensional velocities of 300--500~\kms, with a high velocity
tail extending beyond 1000~\kms.

A variety of physical mechanisms have been proposed as the origin of
high velocities.  Perhaps the first was disruption of binaries through
mass loss in supernovae (\citealt{b61}; \citealt*{ggo70};
\citealt{it96}), although it is difficult for binary disruption alone
to account for some of the highest observed velocities
(\citealt*{hla93}; \citealt{cvb+05}). The most natural source of such
high velocities appears to be asymmetries in the birth supernovae of
pulsars \citep{s70,vdhvp97,pzvdh99}, although exactly how an asymmetry
in the core collapse process in a supernova explosion is converted to
a birth kick imparted to a nascent neutron star remains unclear
\citep*{lcc01}.  While hydrodynamic or convective instabilities are the
most plausible route
\citep[e.g.,][]{bh96,jm96,lg00,spj+04,jsk+05,skjm06,bld+06}, more
exotic mechanisms such as asymmetric neutrino emission in the presence
of strong magnetic fields \citep{al99b} or some combination of the
above \citep{sbhf05} cannot be ruled out.

Of these kick mechanisms, many predict kicks vectors that relate to
the spin axis orientation of the nascent neutron star: the alignment
(or lack thereof) of the natal kick with the neutron star spin axis
could provide a specific discriminant between various mechanisms
(\citealt*{bhf95}; \citealt{sp98,cowsik98,lcc01,romani05}).  Even such
parameters such as the number and timescale of kick components,
coupled with the initial spin period of the neutron star, can be
constrained through observations of an ensemble of sources
(\citealt*{drr99}; \citealt{jhv+06,wlh06,wlh07,nr07,rankin07}).


\subsection{The Crab Pulsar and its Nebula}
The Crab pulsar (\object[PSR B0531+21]{PSR~B0531+21}) and nebula have been observed by
virtually every telescope capable of pointing at the system, and the
Crab may be the most studied system in all of astronomy.  The Crab
nebula possesses a general symmetry axis, visible in images at most
wavelengths.  Recent observations have delineated this axis with
striking clarity \citep[e.g.,][]{hmb+02,nr04}.  The X-ray jet, in
particular, allows us to trace the symmetry axis to the neutron star
location, and provides a natural association with the rotation axis of
the pulsar itself (since  every other vector would be
rotation averaged).  The symmetry axis is roughly aligned with the
proper motion \citep[e.g.,][]{cm99}, but the observational
uncertainties have made the alignment hard to quantify.  A precise proper
motion vector for the Crab pulsar could be quantitatively compared to
the jet direction to establish whether the natal kick is aligned with
the spin axis, as many theories predict.


Given its prominent place in our understanding of neutron stars, it is perhaps
surprising that the proper motion and distance of the Crab pulsar are
not better known.  Compared to many fainter objects, the precision of
our measurements is lacking.  While there were a number of early
attempts to measure the proper motion of the Crab pulsar, these were
generally inconsistent with each other \citep{mink70}.  Perhaps the
first reliable measurement was that of \citet[][hereafter
\citetalias{wm77}]{wm77}, who found\footnote{The analysis of
\citetalias{wm77} was based on the B1950 frame, but at this level of
precision the precession between that and the J2000 frame does not
change the proper motion significantly.}  $(\mu_\alpha,\mu_\delta) =
(-13\pm2,7\pm3)\,\masyr$ from photographic plates spanning
epochs from 1899 to 1976.  There have not been any direct (i.e.\
geometric) distance measurements of the pulsar itself, but the
distance was estimated based on various lines of evidence to lie
between 1.4 and 2.7$\,$kpc \citep{t73}. In spite of the wealth of
observations since then, including a treasure trove of \textit{Hubble
Space Telescope} (\hst) images with a time baseline of $>10$ years,
these estimates had not significantly improved until recently. For
example, \citet[][hereafter \citetalias{cm99}]{cm99} estimate a proper
motion $(\mu_\alpha,\mu_\delta) = (-17\pm3,7\pm3)\,\masyr$,
which is consistent with the earlier estimate, but does not improve
upon its accuracy.

The main obstacle to more accurate measurements is not (as in many
cases) the limitations of faint objects, but rather the fact that the
Crab nebula is too bright for most interferometric radio observations
(it raises the system temperature too much), that there are no
suitable interferometric calibrators nearby, and that the rotational
stability is not sufficient (due to glitches) for precise pulse timing
over a long time baseline.  Because of these reasons, high angular
resolution optical observations are so far the only way to measure the
astrometric parameters (proper motion and parallax) of the Crab
pulsar, and with the current generation of instruments we are limited
to data from \hst.

\citet[][hereafter \citetalias{nr06}]{nr06} attempted a significantly
more detailed astrometric analysis of the Crab pulsar compared to
\citetalias{cm99}, taking advantage of new \hst\ data spanning 7 years
and trying to account for many sources of uncertainty not addressed by
\citetalias{cm99}.  \citetalias{nr06} found a result that is
discrepant with that of \citetalias{cm99}: $(\mu_\alpha,\mu_\delta)
=(-15.0\pm0.8,1.3\pm0.8)\,\masyr$.  This shows a significant
misalignment with the projected spin-axis of the pulsar
($26\degr\pm3\degr$) and as such reverses previously held notions of
spin-kick alignment, but as we discuss below (and as \citetalias{nr06}
acknowledge) even this analysis still is not as accurate as possible.
Additionally, a large number of new observations with the Advanced
Camera for Surveys (ACS) have become publicly available.  These were
taken primarily for studying the dynamics and polarization of the
Crab's pulsar wind nebula \citep[e.g.,][]{hmb+02}, and as such they
were not ideal for astrometry (the exposure times were long enough
that the pulsar saturated, the dithering strategy was not optimal, and
they used a limited range of roll angle), but nonetheless they are an
important resource.

Motivated by the importance of the Crab pulsar in our understanding of
neutron stars and supernova remnants in general, by the large amount
of data available on it, and by the limitations of previous analyses,
we have attempted to re-measure the proper motion of the Crab pulsar
as well as assess the possibility of a parallax measurement with
future data (although given the recent failure of the ACS instrument,
future observations may not be possible until the installation of the
upcoming Wide Field Camera 3).

The organization of our paper is as follows: in \S~\ref{sec:an} we
describe our analysis, noting departures from previous analyses,
although the majority of the fitting techniques are similar to those
we used in \citet*{kvka07}, and we refer readers there for more
details.  After refining the proper motion measurement, we discuss in
\S~\ref{sec:ref} the limitations on our knowledge of the proper motion
imposed by the unknown velocity of the pulsar's progenitor, as well as
uncertainties in the corrections to the pulsar's local standard of
rest.  These transformations and their associated uncertainties limit
the accuracy of the comparison between the proper motion and the
projected spin-axis of the pulsar.  We then give our conclusions in
\S~\ref{sec:conc}.  Finally, we include a discussion of the prospects
for a parallax distance for the Crab pulsar in Appendix~\ref{sec:par}.
In what follows, we define our proper motions in Right Ascension and
Declination $(\mu_\alpha,\mu_\delta)$ so that the scales are the same
and no $\cos\delta$ term is necessary.  All uncertainties are
1-$\sigma$ unless otherwise stated, and all position angles are
measured east of north.

\section{Analysis}
\label{sec:an}
We started our analysis by examining the available archival \hst\ data
for the Crab pulsar.  Twenty observations using the Wide Field and Planetary
Camera 2 (WFPC2) with the F547M filter (a filter centered at $V$-band
but somewhat narrower, designed to avoid bright emission lines)
spanning $\sim$2 years were analyzed by \citetalias{cm99}, who were
able to determine a proper motion for the pulsar that agreed with that
obtained from the ground \citepalias{wm77}. However, the precisions of
both of those measurements were limited and we have a number of
reasons to suspect the analysis of the \hst\ data.

\begin{deluxetable*}{c c c c c c r r c c c}
\tablecaption{Observation Summary\label{tab:obs}}
\tablewidth{0pt}
\tabletypesize{\scriptsize}
\tablehead{
\colhead{Pair} & \colhead{Root\tablenotemark{a}} & \colhead{MJD} & \colhead{Date} &
\colhead{Instrument/} & \colhead{Exp.} &
\multicolumn{2}{c}{Crab $(x,y)_{\rm raw}$\tablenotemark{b}} &
\colhead{PA} &
\colhead{$N_{\rm stars}$\tablenotemark{c}} & \colhead{NR06}\\ \cline{7-8}
\colhead{Number} & \colhead{Name} & & &\colhead{Detector} &&&&&&\colhead{Group\tablenotemark{d}}\\
 & & & & & \colhead{(sec)} & \multicolumn{2}{c}{(pixels)} &\colhead{(deg.)} \\
}
\startdata
\phn1 & \dataset[ADS/Sa.HST#U2BX0501T]{u2bx05} & 49723.7 & 1995-Jan-07 & WFPC2/WF3 & \phn800.0 & 162.98 & 130.19 & \phs309.0 & 11 & 1\\
\phn2 & \dataset[ADS/Sa.HST#U2BX0501T]{u2bx05} & 49723.8 & 1995-Jan-07 & WFPC2/WF3 & 1000.0 & 150.84 & 117.81 & \phs309.0 & 11 & 1\\
\phn3 & \dataset[ADS/Sa.HST#U2U60101T]{u2u601} & 49943.7 & 1995-Aug-15 & WFPC2/PC & 1000.0 & 470.47 & 371.54 & \phn$-$48.6 & \phn5 & \nodata\\
\phn4 & \dataset[ADS/Sa.HST#U2U60201T]{u2u602} & 50026.6 & 1995-Nov-06 & WFPC2/PC & 1000.0 & 340.13 & 300.08 & \phn$-$25.3 & \phn4 & \nodata\\
\phn5 & \dataset[ADS/Sa.HST#U2U60301T]{u2u603} & 50080.4 & 1995-Dec-29 & WFPC2/PC & 1000.0 & 296.63 & 600.35 & \phs128.7 & \phn4 & 2\\
\phn6 & \dataset[ADS/Sa.HST#U2U60401T]{u2u604} & 50102.3 & 1996-Jan-20 & WFPC2/PC & 1000.0 & 297.29 & 600.99 & \phs128.7 & \phn4 & 2\\
\phn7 & \dataset[ADS/Sa.HST#U2U60501T]{u2u605} & 50108.3 & 1996-Jan-26 & WFPC2/PC & 1000.0 & 297.06 & 603.05 & \phs128.7 & \phn4 & 2\\
\phn8 & \dataset[ADS/Sa.HST#U2U60601T]{u2u606} & 50114.5 & 1996-Feb-02 & WFPC2/PC & 1000.0 & 297.83 & 600.07 & \phs128.7 & \phn4 & 2\\
\phn9 & \dataset[ADS/Sa.HST#U2U60701T]{u2u607} & 50135.3 & 1996-Feb-22 & WFPC2/PC & 1000.0 & 297.64 & 601.39 & \phs128.7 & \phn4 & 2\\
10 & \dataset[ADS/Sa.HST#U2U60801T]{u2u608} & 50189.6 & 1996-Apr-17 & WFPC2/PC & 1000.0 & 265.89 & 598.65 & \phs128.7 & \phn4 & 2\\
11 & \dataset[ADS/Sa.HST#U61M0101R]{u61m01} & 51580.1 & 2000-Feb-06 & WFPC2/WF3 & 1100.0 & 372.75 & 273.44 & \phs312.7 & 17 & 3\\
12 & \dataset[ADS/Sa.HST#U61M0201R]{u61m02} & 51589.9 & 2000-Feb-16 & WFPC2/WF3 & 1100.0 & 372.77 & 273.30 & \phs312.7 & 19 & 3\\
13 & \dataset[ADS/Sa.HST#U61M0301R]{u61m03} & 51600.2 & 2000-Feb-26 & WFPC2/WF3 & 1100.0 & 371.49 & 270.99 & \phs312.7 & 19 & 3\\
14 & \dataset[ADS/Sa.HST#U61M0401R]{u61m04} & 51610.5 & 2000-Mar-08 & WFPC2/WF3 & 1100.0 & 372.25 & 272.79 & \phs312.7 & 17 & 3\\
15 & \dataset[ADS/Sa.HST#U61M0501R]{u61m05} & 51620.4 & 2000-Mar-17 & WFPC2/WF3 & 1100.0 & 372.66 & 273.40 & \phs312.7 & 19 & 3\\
16 & \dataset[ADS/Sa.HST#U50V0401R]{u50v04} & 51796.9 & 2000-Sep-10 & WFPC2/WF3 & \phn918.0 & 216.87 & 342.64 & \phs132.7 & \phn8 & 5\\
17 & \dataset[ADS/Sa.HST#U50V0501R]{u50v05} & 51809.0 & 2000-Sep-22 & WFPC2/WF3 & 1090.0 & 216.78 & 342.43 & \phs132.7 & \phn9 & 5\\
18 & \dataset[ADS/Sa.HST#U50V0601R]{u50v06} & 51818.8 & 2000-Oct-02 & WFPC2/WF3 & \phn872.0 & 216.88 & 342.45 & \phs132.7 & \phn8 & 5\\
19 & \dataset[ADS/Sa.HST#U50V0701R]{u50v07} & 51829.9 & 2000-Oct-13 & WFPC2/WF3 & 1200.0 & 217.17 & 341.14 & \phs132.7 & \phn9 & 5\\
20 & \dataset[ADS/Sa.HST#U50V0801R]{u50v08} & 51840.8 & 2000-Oct-24 & WFPC2/WF3 & \phn916.0 & 216.75 & 342.55 & \phs132.7 & \phn7 & 5\\
21 & \dataset[ADS/Sa.HST#U50V1001R]{u50v10} & 51863.0 & 2000-Nov-15 & WFPC2/WF3 & \phn889.5 & 216.78 & 342.48 & \phs132.7 & \phn7 & 5\\
22 & \dataset[ADS/Sa.HST#U50V1101R]{u50v11} & 51873.1 & 2000-Nov-25 & WFPC2/WF3 & 1100.0 & 225.62 & 334.16 & \phs132.7 & \phn8 & 5\\
23 & \dataset[ADS/Sa.HST#U50V1201R]{u50v12} & 51884.4 & 2000-Dec-06 & WFPC2/WF3 & 1000.0 & 216.78 & 340.60 & \phs132.7 & \phn8 & 5\\
24 & \dataset[ADS/Sa.HST#U50V1301R]{u50v13} & 51896.3 & 2000-Dec-18 & WFPC2/WF3 & 1000.0 & 371.97 & 274.06 & \phs312.7 & 17 & 6\\
25 & \dataset[ADS/Sa.HST#U50V1401R]{u50v14} & 51906.7 & 2000-Dec-29 & WFPC2/WF3 & 1000.0 & 371.88 & 273.82 & \phs312.7 & 19 & 6\\
26 & \dataset[ADS/Sa.HST#U50V1501R]{u50v15} & 51918.5 & 2001-Jan-09 & WFPC2/WF3 & 1200.0 & 387.46 & 275.98 & \phs312.7 & 19 & 6\\
27 & \dataset[ADS/Sa.HST#U50V1701R]{u50v17} & 51939.4 & 2001-Jan-30 & WFPC2/WF3 & 1200.0 & 387.68 & 276.10 & \phs312.7 & 18 & 6\\
28 & \dataset[ADS/Sa.HST#U50V1801R]{u50v18} & 51950.5 & 2001-Feb-10 & WFPC2/WF3 & 1200.0 & 383.72 & 278.22 & \phs312.7 & 17 & 6\\
29 & \dataset[ADS/Sa.HST#U50V1901R]{u50v19} & 51961.5 & 2001-Feb-21 & WFPC2/WF3 & 1200.0 & 388.42 & 276.09 & \phs312.7 & 17 & 6\\
30 & \dataset[ADS/Sa.HST#U50V2001R]{u50v20} & 51972.8 & 2001-Mar-05 & WFPC2/WF3 & 1200.0 & 389.02 & 275.64 & \phs312.7 & 18 & 6\\
31 & \dataset[ADS/Sa.HST#U50V2101R]{u50v21} & 51983.3 & 2001-Mar-15 & WFPC2/WF3 & 1200.0 & 389.75 & 275.45 & \phs312.7 & 18 & 6\\
32 & \dataset[ADS/Sa.HST#U50V2201R]{u50v22} & 51994.5 & 2001-Mar-27 & WFPC2/WF3 & 1200.0 & 389.81 & 275.43 & \phs312.7 & 18 & 6\\
33 & \dataset[ADS/Sa.HST#U50V2301R]{u50v23} & 52005.2 & 2001-Apr-06 & WFPC2/WF3 & 1200.0 & 389.98 & 275.55 & \phs312.7 & 17 & 6\\
34 & \dataset[ADS/Sa.HST#U50V2401R]{u50v24} & 52016.3 & 2001-Apr-17 & WFPC2/WF3 & 1200.0 & 389.50 & 275.67 & \phs312.7 & 18 & 6\\
35 & \dataset[ADS/Sa.HST#J8Q410011]{j8q410f} & 52859.8 & 2003-Aug-09 & ACS/WFC1-2K & 1100.0 & 1322.68 & 1047.20 & \phn$-$94.8 & 60 & \nodata\\
36 & \dataset[ADS/Sa.HST#J9FX01011]{j9fx01e} & 53619.7 & 2005-Sep-07 & ACS/WFC1-2K & 1150.0 & 1317.77 & 1044.48 & \phn$-$95.0 & 55 & \nodata\\
37 & \dataset[ADS/Sa.HST#J9FX02011]{j9fx02l} & 53628.8 & 2005-Sep-16 & ACS/WFC1-2K & 1150.0 & 1317.91 & 1044.13 & \phn$-$94.8 & 60 & \nodata\\
38 & \dataset[ADS/Sa.HST#J9FX03011]{j9fx03u} & 53638.7 & 2005-Sep-26 & ACS/WFC1-2K & 1150.0 & 1318.42 & 1043.14 & \phn$-$94.6 & 56 & \nodata\\
39 & \dataset[ADS/Sa.HST#J9FX04011]{j9fx04z} & 53645.7 & 2005-Oct-03 & ACS/WFC1-2K & 1150.0 & 1318.45 & 1042.15 & \phn$-$94.4 & 60 & \nodata\\
40 & \dataset[ADS/Sa.HST#J9FX05011]{j9fx05l} & 53655.7 & 2005-Oct-13 & ACS/WFC1-2K & \phn975.0 & 1318.91 & 1041.79 & \phn$-$94.2 & 60 & \nodata\\
41 & \dataset[ADS/Sa.HST#J9FX06011]{j9fx06s} & 53665.7 & 2005-Oct-23 & ACS/WFC1-2K & 1150.0 & 1319.41 & 1039.97 & \phn$-$93.9 & 58 & \nodata\\
42 & \dataset[ADS/Sa.HST#J9FX07011]{j9fx07x} & 53673.7 & 2005-Oct-31 & ACS/WFC1-2K & 1150.0 & 1318.87 & 1038.25 & \phn$-$93.6 & 52 & \nodata\\
43 & \dataset[ADS/Sa.HST#J9FX08011]{j9fx08f} & 53682.3 & 2005-Nov-08 & ACS/WFC1-2K & 1150.0 & 1318.60 & 1036.49 & \phn$-$93.2 & 55 & \nodata\\
44 & \dataset[ADS/Sa.HST#J9FX09011]{j9fx09j} & 53690.7 & 2005-Nov-17 & ACS/WFC1-2K & 1150.0 & 1319.01 & 1033.36 & \phn$-$92.6 & 56 & \nodata\\
45 & \dataset[ADS/Sa.HST#J9FX10011]{j9fx10u} & 53699.7 & 2005-Nov-26 & ACS/WFC1-2K & 1150.0 & 1318.17 & 1027.05 & \phn$-$91.6 & 53 & \nodata\\
46 & \dataset[ADS/Sa.HST#J9FX11011]{j9fx11c} & 53709.5 & 2005-Dec-05 & ACS/WFC1-2K & 1150.0 & 1281.89 & 871.96 & \phn$-$62.2 & 41 & \nodata\\
47 & \dataset[ADS/Sa.HST#J9FX21011]{j9fx12h} & 53718.6 & 2005-Dec-15 & ACS/WFC1-2K & 1150.0 & 1269.60 & 851.69 & \phn$-$57.2 & 48 & \nodata\\

\enddata
\tablenotetext{a}{Root name of the dataset in the STScI archive.}
\tablenotetext{b}{Raw pixel position of the Crab pulsar.}
\tablenotetext{c}{Number of reference stars that we used on each
  image, excluding the Crab pulsar.}
\tablenotetext{d}{Group number assigned by \citetalias{nr06}.}
\tablecomments{Each pair consists of two identical observations taken
  for cosmic-ray rejection (\texttt{CRSPLIT}$=2$).  All of the WFPC2
  observations were taken with the F547M filter, while the ACS
  observations were taken with the F550M filter.  We processed each of
  the exposures separately.}
\end{deluxetable*}

When we examined the \hst\ data used in the prior analyses in detail
we noticed that the pulsar itself was very saturated.  This is not
unexpected: using the WFPC2 exposure-time calculator (ETC) and
\texttt{synphot}\footnote{See
\url{http://www.stsci.edu/resources/software\_hardware/stsdas/synphot}.},
with the input spectrum from \citet{sll+00}, we would expect about
$600\mbox{\,counts\,s}^{-1}$, or 600,000 total counts from the pulsar in
a typical 1000-s exposure.  Even with a gain of $15\mbox{\,e}^{-}/\mbox{ADU}$ (used in some of the observations, where ADU is
analog-digital units), this is still 40,000~ADU over just a few
pixels, and WFPC2 saturates near 3500~ADU (depending on the gain).
The pulsar is saturated in all of the F547M data by up to a factor of
10 in both PC and WF observations, but this is not mentioned by
\citetalias{cm99}, although they do check for saturation among the
reference stars.  Saturation can severely degrade WFPC2 data since the
point-spread function (PSF) is undersampled by the detector and most
of the flux is concentrated in a small number of pixels.  Measuring
positions is particularly difficult for saturated stars, since most of
the positional information in normal cases comes from the central
pixels where the PSF is changing most rapidly.  When these central
pixels are saturated, one is forced to fit using the gently sloping
wings of the PSF, and they provide a very weak handle on the position.
In the next section, we discuss in detail how much the saturation
degrades our astrometric accuracy.  In addition, there is no dithering
between the exposures taken at the same epoch; such dithering can help
overcome the undersampling of WFPC2 (see \citealt{ak00}).

The analysis done by \citetalias{nr06} improved upon that of
\citetalias{cm99}.  \citetalias{nr06} did not resample the data
(resampling can degrade the astrometry and introduce numerical
artifacts) but treated each position measurement
individually.  Additionally, they used 15 reference stars instead of
four, used an improved distortion solution, fit for the proper motions
of reference stars and for the orientations and scales of the
exposures, and attempted to account for the saturation of the Crab pulsar.
Finally, they used many more exposures.  However, this approach was
still not ideal, as mentioned by \citetalias{nr06} themselves, as they
used Gaussian fits instead of effective PSF (ePSF; \citealt{ak00})
fits for the position measurements.  Overall, including estimates of
the uncertainties due to residual distortion error, \citetalias{nr06}
find  uncertainties of $\pm0.8\,\masyr$ in each
coordinate of the proper motion.

We have attempted to improve on the analyses of both \citetalias{cm99} and
\citetalias{nr06}.  The most significant improvement comes from using
many more observations: in addition to the large number of WFPC2
observations used by \citetalias{nr06}, we used a sizable number of
exposures with the ACS/Wide Field Camera (ACS/WFC), a small number of
which were discussed by \citetalias{nr06} but not incorporated into
their final analysis.  We also use proper ePSF measurements and the
latest distortion solutions.  

As we show, our analysis yields measurement uncertainties on the
proper motion of $\pm0.4\,\masyr$.  At this level, we must
consider in detail the reference frame of the measurement that we make
and the corrections necessary to transform our measurement into the
reference frame of the Crab nebula.   We do this in \S~\ref{sec:ref},
and find that the reference frame uncertainties dominate the
measurement uncertainties by a wide margin.

From the WFPC2 observations, we selected only those taken with the
F547M filter: there were too few observations with the other filters
($\leq 6$ in a given filter) to allow us to properly characterize the
data.  We also restricted our data to observations where the Crab
pulsar was either on the Planetary Camera chip (PC) or Wide Field chip
\#3 (WF3) and only analyzed the chip that the pulsar was on, as for
the other data sets either the pulsar was too close to the central
reflecting pyramid\footnote{See
\url{http://www.stsci.edu/instruments/wfpc2/Wfpc2\_dhb/wfpc2\_ch1.html}.}
for reliable astrometry (pixel values $x$ or $y<100$, generally) or
there were again too few observations for proper characterization.
We included only the ACS observations taken with the F550M filter
(similar to the F547M filter), as the data with other filters were
either too sparse to characterize or had no reliable distortion
solutions/point spread functions (e.g., data taken through
polarizers).  These observations used the \texttt{WFC1-2K} mode, where
only one of the two WFC detectors is active, and only a
$2048\times2048\,$pixel sub-region of that detector is read out (of the
complete $4096\times2048\,$pixel detector), thus giving a field-of-view
that is one quarter the area of the complete ACS/WFC.  Our final set of
observations is listed in Table~\ref{tab:obs}, and consists of 47
pairs of exposures, where each pair consists of two exposures at the
same position taken for cosmic-ray rejection (i.e.\
\texttt{CRSPLIT}$=2$).

We took the pipeline processed images from the \hst\ archive, leaving
them at the flatfielded stage but not applying any drizzling
\citep{kfhh02}.  We identified cosmic rays from the \texttt{CRSPLIT}
pairs, using the task \texttt{driz\_cr} from the STSDAS dither package
\citep{fm98} for the WFPC2 data and using the pipeline-produced
data-quality extensions for the ACS data.  For each individual
exposure (we did not combine \texttt{CRSPLIT} pairs or different
WFPC2 detectors), we performed ePSF astrometry, using the distortion
solutions and ePSFs for WFPC2 from \citet{ak03} and for the ACS/WFC
from \citet{ak06}.  We note that the pulsar was substantially
saturated on all of the exposures, both ACS and WFPC2, as were a small
number of other stars. For the WFPC2 data, where the undersampling is
particularly bad, we fit the saturated stars with a larger ePSF
(6-pixel radius) that extends further into the wings, which
significantly improved the reliability of the measurements in tests
that we did.  For the ACS data the effects of saturation were not as
bad, as the instrument has a larger dynamic range and the better
spatial sampling means that more unsaturated pixels are available in
the wings of the PSF.  Therefore we applied the standard \citet{ak06}
ePSF technique to the ACS observations of saturated sources.  To avoid
measurements that were contaminated by cosmic rays, we rejected
individual star positions of non-saturated sources if there was even
one pixel contaminated by a cosmic ray (identified by the algorithms
above) within the central $5\times5\,$pixel box used for the
astrometry; however, we did not reject measurements of saturated
sources (including the pulsar), since the cosmic ray identification
routines could not distinguish between real cosmic rays and the
effects of saturation.  We also rejected all WFPC2 measurements with
$x$ or $y<100$ to avoid the effects of the central reflecting pyramid.

We assembled all of the position measurements, starting with the ACS
observations.  These had a $100\arcsec\times100\arcsec$ field of view
and we were able to identify up to 73 stars besides the pulsar on
those images, but we rejected 7 of them as they were too close to the
edges.  This left us with 66 unique stars, of which we detected up to
60 on any individual image.  We then identified those stars on the
WF3 and PC exposures and found that there were an additional 8 stars that we
could identify that were not on the ACS images, so we have a total of
74 stars that we used.  Note that we did not use star \#0 from
\citetalias{nr06} as it was too saturated 
to have reliable measurements, but we have a sufficiently large number
of other stars that our analysis is still robust.  Also, the preferred
solution from \citetalias{nr06} used groups 1, 3, and 6 in their
numbering; their group 1 corresponds to our pairs 1 and 2, their group
3 corresponds to our pairs 11--15, and their group 6 corresponds to
our pairs 24--33 (see Tab.~\ref{tab:obs}); \citetalias{cm99} used data
from \citetalias{nr06}'s group 3 as well as two observations where the
pulsar was on the WF2 detector, but like \citetalias{nr06} we chose
not to analyze those observations since there were too few to
understand the uncertainties (see below).

\subsection{Measurement Uncertainty Estimation}
\label{sec:unc}
To properly combine all of the position measurements in a
statistically meaningful analysis, we need estimates of the individual
astrometric uncertainties.  We took advantage of the \texttt{CRSPLIT}
pairs, between which there should be at most a very small
shift/transformation due to telescope jitter (variations in telescope
pointing) and breathing (variations in the detector scale due to
thermal fluctuations; e.g., \citealt{ak06}).  For each instrument
separately (ACS/WFC, WFPC2/PC, and WFPC2/WF3) we compared the position
of each star with that in the other \texttt{CRSPLIT} image, measuring
position differences $\Delta x_{s,p}$ and $\Delta y_{s,p}$, where $s$
is an index that runs over the number of stars and $p$ is an index
that runs over the number of exposure pairs $N_p$.  We then
determined for each star the variance of those position
differences:
\begin{equation}
\sigma_{x,s}^2=\frac{1}{N_p-1}\sum_{p=1}^{N_p}\Delta x_{s,p}^2
\end{equation}
and the same for $\sigma_{y,s}$.  Without dithering, there could be
additional uncertainties due to pixel-phase errors (errors from stars
landing at different positions within a pixel; \citealt{ak00}) or from uncorrected
distortion that we do not see from this \texttt{CRSPLIT} analysis, but
as we see later our estimated uncertainties were largely sufficient.

\begin{figure}
\plotone{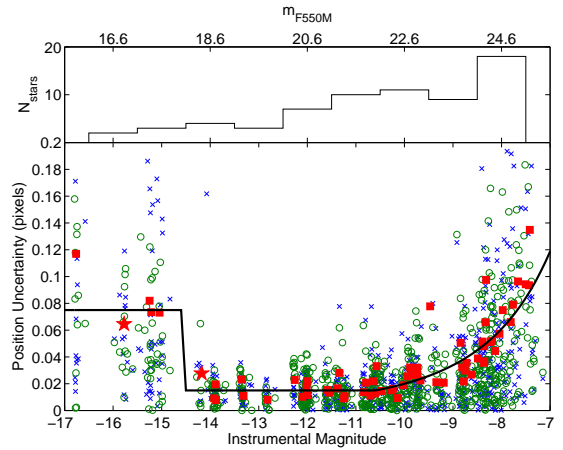}
\caption{Bottom panel: Position uncertainty for each star as a
  function of instrumental magnitude ($-2.5\log_{10}\mbox{Counts}$ in
  a $5\times5\,$pixel box) for the ACS/WFC data.  The crosses are the
  $x$ position residuals, and open circles are the $y$ residuals, and
  the filled squares are the standard deviations of $x$ and $y$
  combined, all divided by $\sqrt{2}$ so as to be appropriate for a
  single observation.  The solid line is the uncertainty model derived
  in \S~\ref{sec:unc}.  The Crab pulsar is the object at instrumental
  magnitude of $-15.8$ and our primary reference star (\#4) is the
  object at $-14.1$, both with a red stars. Top panel: number of stars
  used as a function of $m_{\rm F550M}$, where we have taken $m_{\rm
  F550M}=m_{\rm inst}+32.6$.  The abscissae of the two panels are
  aligned.}
\label{fig:uncertwfc}
\end{figure}

For the ACS/WFC data, we started with the relation of positional
uncertainty as a function of instrumental magnitude
($-2.5\log_{10}\mbox{Counts}$ in a $5\times5\,$pixel box) for the WFC
from \citet[][Fig.~13]{ak06}.  Since the brightest non-saturated stars
will have more uncertain measurements than those in \citet{ak06} and
we have not derived an updated ePSF or distortion solution, we had an
artificial minimum at $0.015\,$pixels for bright stars.  To account
for saturation, which occurs at $m_{\rm inst}<-14$ for the WFC, we
increased the uncertainty to 0.075$\,$pixels.  This trend gives a
reasonably good fit to the measured standard deviations (divided by
$\sqrt{2}$), as shown in Figure~\ref{fig:uncertwfc}, and
(\S~\ref{sec:pm}) also works well for the final analysis.  We used
several saturated stars in addition to the Crab pulsar: they
contribute very little to the actual fit, but by examining their
residuals (e.g., Fig.~\ref{fig:uncertwfc}) we gain a check on how well
we can expect the pulsar data to fit.  On the faint end, we included
stars down to a WFC instrumental magnitude of $-7.5$: we could have
chosen a brighter limit with smaller uncertainties, but as shown in
the top panel of Figure~\ref{fig:uncertwfc} the number of stars is
increasing and this increases the reliability of the fit.  This is
especially true for the WFPC2 data, where there are fewer reference
stars overall.

For the WFPC2 data, we started with the trend found in \citet*{kvka02},
and we used the same trend for uncertainty in pixels as a function of
magnitude  for both the PC and WF3.  We then followed the procedure
outlined above, although we found that we had to multiply the trend
for the WF3 data by a factor\footnote{Since the trend was derived for
the PC it is not surprising that its absolute scale should be
different for the WF chips, but the shape seems consistent.} of 1.5.
For the saturated stars, which had $m_{\rm inst}<-10$, we increased
the uncertainty to 0.1$\,$pixel for the PC and 0.15$\,$pixel for WF3.

For both the ACS and WFPC2 data, we used the fits for the uncertainties as a
function of magnitude rather than the actual uncertainty for each star
as the measured uncertainty is estimated from only a few measurements
and is therefore noisy, while the fit is more predictable.
\citet{kvka07} experimented in detail with different types of
uncertainties and found that they largely did not affect the final
results.

\subsection{Near-IR Color-Magnitude Diagram}
\label{sec:cmd}
The interpretation of our results depends critically  on the distances of
the field stars to which we reference the Crab pulsar's proper motion.
We have thus derived a color-magnitude diagram of stars near the
pulsar, using near-IR photometric observations
with the Wide Field Infrared Camera (WIRC; \citealt{weh+03}) on the
Palomar 200-inch telescope.  The observations were on
2003~November~19, and we exposed for $15\times20\,$s in the $J$ and
$K_s$ filters.  The seeing was not very good, and averaged $1\farcs5$.
For the reduction, we subtracted dark frames, then produced a sky
frame for subtraction by taking a sliding box-car window of 4
exposures on either side of a reference exposure.  We then added the
exposures together, identified all the stars, and produced masks for
the stars that were used to improve the sky frames in a second round
of sky subtraction.  We referenced the astrometry and photometry to
the Two Micron All Sky Survey (2MASS; \citealt{2mass}), using $>200$
well-detected stars that were not knots of nebulosity.

\begin{figure}
\plotone{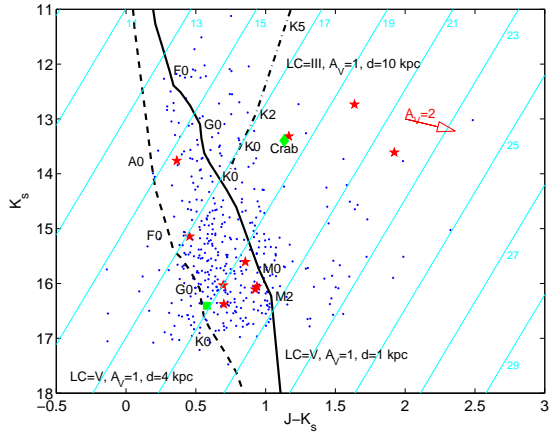}
\caption{Near-IR color-magnitude diagram of the Crab pulsar field.  We
  plot $K_s$ magnitude vs.\ $J-K_s$ color for the stars detected on
  our $8\arcmin\times8\arcmin$ WIRC image.  The uncertainties are
  dominated by uncertainties in our photometric zeropoints of $\approx
  0.03\,$mag.  The tracks are from \citet{allen}: the solid track is
  the main sequence (luminosity class V) for $A_V=1.0$ and distance
  1$\,$kpc, the dashed track is the main sequence for $A_V=1.0$ and
  $d=4\,$kpc, and the dot-dashed track is the giant branch (luminosity
  class III) for $A_V=1.0$ and $d=10\,$kpc; some stellar types are
  labeled.  We also show a reddening vector for $A_V=2.0$.  The
  diagonal lines are approximate lines of constant $V$ (or F550M) magnitude,
  assuming $V-K_{\rm s}\approx 4.3(J-K_{\rm s})$.  Note that $V$
  magnitude is very close to both $m_{\rm F547M}$ and $m_{\rm
  F550M}$. The stars are the 10 objects used in our \hst\ astrometry
  that we could detect in the WIRC images, the Crab pulsar is the
  green diamond (labeled), and star \#4 from \citetalias{nr06} (which is
  our proper motion reference star) is the green square.
\label{fig:cmd}}
\end{figure}

We used \texttt{sextractor} \citep{ba96} to produce a source list for
the $K_s$-band image, and then ran \texttt{sextractor} again on the
$J$-band image using the $K_s$-band source list as a set of starting
positions.  Finally, we produced the color-magnitude diagram shown in
Figure~\ref{fig:cmd}.  

Given the depth of the images and the poor seeing, we could not
measure near-IR magnitudes for most of the stars that we used for
astrometry.  These stars were mostly within the Crab nebula and the
high background limited the ground-based image.  Therefore our
color-magnitude diagram predominantly includes stars from the full
$8\arcmin\times8\arcmin$ field outside the Crab nebula.  We do not
expect that the different astrometric and photometric samples will be
biased relative to one another, except that the limiting magnitude in
the near-IR is brighter: the photometric stars are those that are
either very bright and/or are outside the Crab nebula, while the
astrometric stars include those inside the nebula for a range of
brightnesses.  We estimated a rough conversion between our near-IR
photometry and our \hst\ photometry using \texttt{synphot} for
$A_V=1.0\,$mag, where we find $V-K_{\rm s}\approx 4.3(J-K_{\rm s})$ and
$V\approx m_{\rm F550M}\approx m_{\rm F547M}$ (since the STMAG system
is based on the Vega system at $V$, no zeropoint offset is necessary).
A rough photometric calibration for the \hst\ data can be done with
the instrumental magnitudes that we measure, such that $m_{\rm
F550M}\approx m_{\rm inst}+32.6$, where we neglect any aperture
corrections.  As can be seen from Figure~\ref{fig:uncertwfc}, the
majority of the stars are at $m_{\rm F550M}\gsim 21$, which is fainter
than most stars in Figure~\ref{fig:cmd} but does not imply drastically
different distances.  The WFC saturation limit of $m_{\rm inst}=-14$
translates to $m_{\rm F550M}\approx 19$, and indeed the stars above
this line in Figure~\ref{fig:cmd} are saturated.

From this we see that the majority of the stars in this field are
consistent with main-sequence stars at distances of 1--4$\,$kpc; a few
may be more distant giants, and some of the astrometric reference
stars may be among these, but they could also be slightly more
reddened main-sequence stars (note that we are not accounting for the
effects of metallicity).  There should not be many stars that are much
closer, as they would have to be rather redder than the more distant
stars that we measure.

\subsection{Fitting for the Astrometric Parameters}
\label{sec:pm}
\begin{figure*}
\epsscale{.8}
\plotone{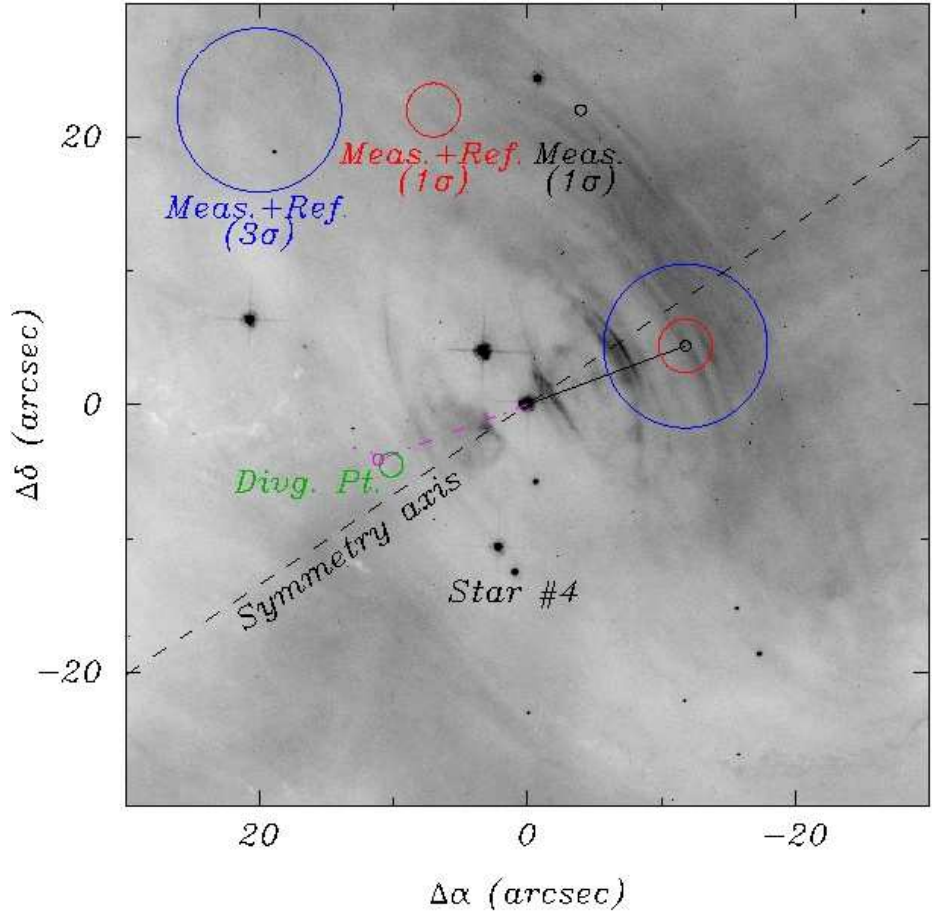}
\caption{Proper motion of the Crab pulsar, shown on a drizzled
  \citep{kfhh02} ACS/WFC image (pair 35).  The vector indicated by the
  solid line shows the proper motion of the pulsar, corrected according
  to \S~\ref{sec:ref} for the effects of differential Galactic
  rotation and solar motion, and  showing the position of the pulsar 1000~yrs
  from now.  The inner-most circle is the 1-$\sigma$ statistical
  uncertainty; the middle circle is the 1-$\sigma$ combined
  (measurement error plus correction for DGR and LSR plus inclusion of
  uncertainty in the progenitor's peculiar motion relative to its
  local standard of rest) uncertainty; and the outer circle the is
  3-$\sigma$ combined uncertainty.  We also label our primary
  reference star (\#4 from \citetalias{nr06}) and the nebula's
  symmetry axis (at position angle $304\degr$ north through east;
  \citealt{nr04}) which we take to be the projected rotation axis of
  the pulsar.  The magenta dot-dashed line indicates our proper motion
  projected back to $1054$~CE and the green circle indicates the
  divergent point (with $\pm 1$-$\sigma$ uncertainties) found by
  \citetalias{wm77} for the Crab's filaments.  North is up, and east
  to the left.}
\label{fig:wfc}

\end{figure*}

Once we have our data set with uncertainties as derived in
\S~\ref{sec:unc}, we are in a position to fit for the positions and proper
motions of each star, as well as the plate-scales, orientations, and
central position of each exposure (see \citealt{kvka07} for a detailed
description of the fitting procedure).  For the fitting, we need to
set the absolute plate-scale and orientation, which we did by assuming
the plate-scale and orientation from the first exposure from pair~35
are the nominal values of $50\,\masyr$ and the rotation
from the image header; this exposure is no more likely than any other
to have the correct plate-scale or orientation, but that will just
lead to absolute uncertainties on the proper motion of
$0.1$\% or less\footnote{We have verified this by fitting our final reference
positions to positions derived from the WIRC data, which are
referenced to 2MASS, which is tied to the International Coordinate
Reference System.  We find rotations of $<0.1\degr$ and scale changes
of $<0.1$\%.  Also see \S~\ref{sec:absast} and \citet{vdmac+07}.}.
We then also assumed that the orientation of the first exposure from
pair 36 is the header value, since without two exposures with known
orientations the fitting procedure could lead to a net rotation of the
data with time that is compensated by a bulk proper motion.  The
fitting process might also compensate for a secularly increasing
offset between exposures by introducing a fictitious bulk proper
motion for the ensemble of stars. In order to prevent this, we
initially assigned a star to have zero proper motion. This does not
actually define the reference frame, since we can arbitrarily shift
the proper motions of all of the stars.  For this fixed star we chose
\#4 from \citetalias{nr06}: this star has the advantage that it is
close to the Crab pulsar, so it is on almost all exposures (73 of 94,
after rejecting individual measurements for cosmic rays as described
above) and is bright but not saturated (see Fig.~\ref{fig:wfc}).  From
\S~\ref{sec:cmd} and Figure~\ref{fig:cmd} star \#4 appears to be roughly
at $\sim 4\,$kpc.

Once we have measured source positions and estimated uncertainties at
each epoch, we directly use the different observations to measure a
proper motion. We fit simultaneously for the positions and proper
motions of each star (with the proper motion of \#4 fixed to zero) and
the transformation parameters for each exposure (with the rotations
and plate-scales fixed as discussed above for the first exposures in
pairs~35 and 36), including all of the WFPC2 and ACS data in the fit.
Each exposure had a 6-parameter transformation, such as that used by
\citet{kvka07} and \citet{ak06}.  Such a transformation is able to
deal implicitly with linear variations in the distortion caused by
breathing or systematic effects \citep{anderson07}.  For the ACS data
the fits give scale uncertainties of $\approx 0.0005$\% and position
angle uncertainties of $0\fdg0003$, and shift
uncertainties of $<0.01\,$pixel.  For the WFPC2 data the results are
somewhat worse, largely due to the smaller number of stars: WF3 has
scale uncertainties of $\approx 0.02$\% and position angle uncertainties of
$0\fdg005$, and shift uncertainties of 0.02--0.1$\,$pixel (depending on
the number of stars included), while the PC has
scale uncertainties of $\approx 0.07$\%, rotation uncertainties of
$0\fdg05$, and shift uncertainties of $0.1\,$pixel.


Initially we achieved a reasonable fit, with $\chi^2=5935.9$ for 3808
degrees of freedom (dof; we had 2322 observations of both $x$ and $y$
for 4644 data points, and 836 free parameters), or $\chi^2_{\rm
red}=1.559$.  This $\chi^2$ results from comparing the computed
positions of every star at every epoch (based on the fitted reference
positions, proper motions, and frame transformations) to the measured
positions; see \citet{kvka07}, Eqn.~A6. However, there were
anomalously large contributions to the total $\chi^2$ from a few
deviant data points.  Beyond the cosmic-ray rejections discussed
above, we rejected an additional 8 measurements that deviated by more
than $10\sigma$ from the best-fit model, and reduced $\chi^2$ to
4473.7 for 3792 dof ($\chi^2_{\rm red}=1.180$; the proper motion of
the pulsar changed by $\ll 1\sigma$ after the rejections).  The
rejected measurements were distributed among the stars and exposures,
and likely represented statistical fluctuations or undetected
cosmic-rays.  After the additional rejections, the fit looked good
overall, with no individual star or exposure dominating the fit.  Note
that the pulsar is saturated, and so its astrometric position
uncertainty is significantly higher than most of the reference
stars. As such, it does not dominate the overall fit.  We tested using
other reference stars and exposures (both ACS and WFPC2 observations),
and the results did not depend on those choices except for a net shift
in the proper motion, but we correct for this below.


As discussed above, all of the proper motions that we fit for are
relative to that of star \#4 from \citetalias{nr06}, but we of course
do not know what the proper motion of star \#4 is, and it does not
make a useful reference frame: with a transverse velocity
dispersion of $\sim 20\,\kms$ and a distance of a few kpc, the proper
motions of random stars are $\sim 1\,\masyr$.  We must
therefore try to determine a reference frame for our measurements in
which the projected motion of the pulsar can be compared sensibly with
the projected orientation of the nebular symmetry axis.  We do so in two steps.
First, we determine the average proper motion of the ensemble of
reference stars.  Next, we consider how those stars are moving
relative to the Sun.

\begin{figure}
\epsscale{1}
\plotone{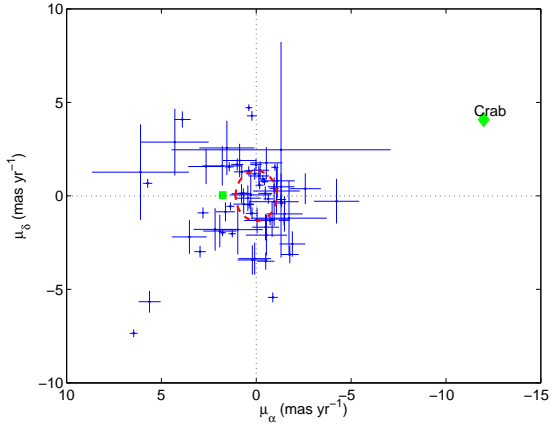}
\caption{Proper motions for each star, shifted so that the net proper
  motion is zero (represented by the dotted lines).  The Crab (green 
  diamond) and star \#4 from \citetalias{nr06} (green square) are
  identified.  The red ellipse shows the standard deviation of the
  reference stars (after rejecting outliers), and the magnitude of the
  shift is shown by the resulting proper motion of star \#4 (whose
  proper motion was fixed to zero during the fit).
\label{fig:bulkpm}
}
\end{figure} 

In Figure~\ref{fig:bulkpm} we plot all of the proper motions that we
measure.  The Crab pulsar clearly has a much larger and more
significant proper motion than the other objects, although we find a
number of other stars with $>3\sigma$ detections of proper motion.
In that figure we have determined the mean proper motion of all of the
stars excluding the Crab pulsar (iteratively rejecting outliers) and
shifted the proper motions to have zero mean.  This shift has a
magnitude of $(\Delta \mu_{\alpha},\Delta
\mu_{\delta})=(-1.8\pm0.2,-0.0\pm0.2)\,\masyr$, moving star
\#4 to its position away from the origin.  The circle in
Figure~\ref{fig:bulkpm} shows the standard deviation of the proper
motions, and it is comparable to the magnitude of the shift.  The
uncertainty in the shift is much smaller, though, since it is the
standard deviation divided by the square root of the number of stars
used (here 49), and the shift in Right Ascension, at least, is statistically
significant.

To see how our results depended on the choices of reference
star/epochs, and on the datasets we used, we made a number of
different fits.  Using the whole ACS+WFPC2 dataset we iterated among a
variety of reference stars, choosing ones that were relatively bright
but not saturated and were close to the pulsar.  We also used
different ACS reference exposures and found that the proper motion
values (corrected to have zero net stellar proper motion) changed by
at most $0.6\,\masyr$, consistent with our uncertainties
given below in Equation~\ref{eqn:pm}.

\begin{deluxetable*}{l l l l@{$\pm$}l l@{$\pm$}l}
\tablecaption{Measured Proper Motions for the Crab Pulsar\label{tab:pms}}
\tablewidth{0pt}
\tablehead{
\colhead{Reference} & \colhead{Data Source} & \colhead{Data Subsets} &
\multicolumn{2}{c}{$\mu_{\alpha}$} & \multicolumn{2}{c}{$\mu_\delta$} \\ \cline{4-7}
 & & & \multicolumn{4}{c}{$(\masyr)$} \\
}
\startdata
\citetalias{wm77} & Plates & \mc{1}{c}{\nodata} & $-13$ &2 & $+7$& 3\\
\citetalias{cm99} & Limited \hst\ WFPC2 & \mc{1}{c}{\nodata} & $-17$& 3 &+7 & 3 \\
\citetalias{nr06} & \hst\ WFPC2 & \citetalias{nr06} groups 1, 3, 6 & $-15$ &0.8 & +1.3& 0.8\\
\citetalias{nr06} & \hst\ WFPC2 & \citetalias{nr06} groups 3, 6 & $-10.9$ & 2.2 & +1.0& 2.0\\
This work & \hst\ WFPC2 & All PC+WF3 & $-15.6$ & 1.8 & $+3.7$ & 1.7\\
This work & \hst\ WFPC2 & \citetalias{nr06} groups 1, 3, 6 &$-13.0$ &
2.4 & +4.6 & 2.3 \\
This work & \hst\ ACS &\mc{1}{c}{\nodata} & \phn$-9.8$ & 1.7 & +4.2 & 1.8\\
This work & \hst\ ACS+WFPC2 &All PC & $-12.5$ & 0.6 & $+4.1$ &0.6\\
This work & \hst\ ACS+WFPC2 &All WF3 & $-12.1$ & 0.6 & $+4.5$ & 0.6\\
This work & \hst\ ACS+WFPC2 &No faint stars\tablenotemark{a} & $-12.0$ & 0.5 & $+3.9$ & 0.5\\
This work & \hst\ ACS+WFPC2 &No saturated stars\tablenotemark{b} &
$-12.3$ & 0.4 & +4.2 & 0.4 \\
This work & \hst\ ACS+WFPC2 & \mc{1}{c}{\nodata} & $-12.0$ & 0.4 & $+4.1$ & 0.4\\
\enddata
\tablecomments{All proper motions are the  values from fitting
  before 
  correction for Solar motion or Galactic rotation. Proper motions from
  this work and \citetalias{wm77} are explicitly in a reference frame
  defined by the background stars in this field; 
  \citetalias{cm99} explicitly assumes that the stars have zero proper
  motion.  Also see
Figure~\ref{fig:allpms}.}
\tablenotetext{a}{We rejected the 27 stars with $m_{\rm
    inst,F550M}>-9.5$ (the upward part of the trend in Fig.~\ref{fig:uncertwfc}).}
\tablenotetext{b}{We rejected the four saturated reference stars.
  This also required removing several of the PC epochs since they had
  too few stars for proper solutions.}
\end{deluxetable*}

\begin{figure}
\plotone{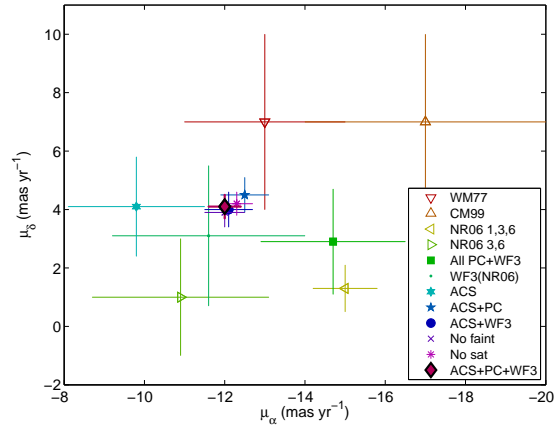}
\caption{A plot of all of the measured proper motions for the Crab
  pulsar from this work and the literature, as given in
  Table~\ref{tab:pms}; these values are the direct fitted results,
  before correction for Solar motion or Galactic rotation.  The
  unfilled triangles (at various orientations) are those values from
  the literature, while the other symbols are our values using various
  subsets of the data [the ``WF3(NR06)'' value is our measurement
  using just the WF3 data used by \citetalias{nr06}].  The best
  measurement is the ``ACS+PC+WF3'' value, but all of our measurements
  are consistent with each other.  }
\label{fig:allpms}
\end{figure}

We then tried to choose different data sets, restricting ourselves to
only the ACS data, only the WFPC2 data, only the WFPC2 data used by
\citetalias{nr06}, and some other combinations.  In particular, we fit
using none of the stars with instrumental magnitudes (from the ACS
data) fainter than $-9.5$: this excludes the portion of
Figure~\ref{fig:uncertwfc} where the trend of uncertainty vs.\
magnitude climbs upwards.  We also fit excluding all of the saturated
stars from the ACS data, with the exception of the Crab pulsar (of
course): this fit required us to remove some of the PC epochs, as
without the saturated stars there were too few reference stars for a
constrained fit.  Such fits proceeded in exactly the same manner as
the fit described above, with the only difference being that we used
different observations as the initial reference; no other special
manipulation was required for these fits.  We found that the corrected
proper motions from all of these data sets were consistent with each
other, as listed in Table~\ref{tab:pms} and shown in
Figure~\ref{fig:allpms}, but were not necessarily consistent with the
values from the literature.  Our result using just the data from
\citetalias{nr06} was consistent with their result, although it had
significantly higher uncertainties.  This is because we took an
uncertainty of 0.15$\,$pix for the WF3 observations of the pulsar, which
is 15$\,$mas, while Figure~3 of \citetalias{nr06} shows their individual
data points as having uncertainties of 5--8$\,$mas.  In contrast, our
measurements using a longer time baseline and with the
higher-precision ACS data were not consistent with the
\citetalias{nr06} values.  In all of these fits the resulting $\chi^2$
values were close to 1.0, indicating that our uncertainty estimates
(\S~\ref{sec:unc}) --- which we derived just by comparing pairs of
observations --- were reasonable for the dataset as a whole, and gave
good values for the different instruments.  The fits that excluded the
faint or the saturated reference stars could be used as our
``default'' fits, but we chose to retain as many stars as possible.
With the exception of the PC epochs, where there are few stars, the
saturated stars largely ride along with the fit and contribute very
little, but they allow us to examine the goodness-of-fit for saturated
objects besides the Crab pulsar, and indeed we find that the data fit
them reasonably well (reduced $\chi^2$ of 0.85 to 1.48).  The fainter
stars contribute more to the fit, but are not dominant, and they allow
us to examine the quality of the fit for a larger number of objects
(again, it is good).

\begin{figure}
\epsscale{.8}
\plotone{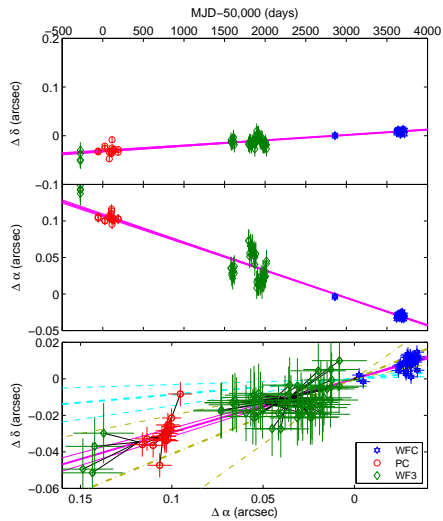}
\caption{Proper motion of the Crab pulsar.  In the top two panels we
  show the position in Declination (top) and Right Ascension (middle),
  plotted against MJD; $\Delta \alpha=0$ and $\Delta \delta=0$
  correspond to our reference epoch of pair 35 (MJD~52859.8).  The
  observations with ACS/WFC are the blue stars, those with WFPC2/PC
  are red circles, and those with WFPC2/WF3 are green diamonds (as
  labeled).  The magenta lines show our best-fit proper motion along
  with $\pm 1\sigma$ uncertainties.  The bottom panel shows the
  position in Right Ascension versus that in Declination: black lines
  connect each observation to the point on the best-fit proper motion
  line at the time of that observation.  We also plot the best-fit
  proper motions along with $\pm 1\sigma$ uncertainties for
  \citetalias{cm99} (dot-dashed lines) and \citetalias{nr06} (dashed
  lines); all proper motions are the final fitted results before
  correction for Solar motion or Galactic rotation.
\label{fig:crabpm}
}
\end{figure}

Using all of the data and our standard choices for reference
exposures/star, the proper motion for just the Crab pulsar was a good
fit, with $\chi^2_{\rm red}=217.3/184=1.18$, although this $\chi^2$ is
not formally correct as the parameters for the exposures are not
properly counted.  We show the astrometry in Figure~\ref{fig:crabpm}.
From this one can see that some pairs of exposures seem to deviate
systematically from the overall trend, such as some of the WFPC2 data
from near MJD~51,800. These are likely related to the changes in
position angle of the observations (Tab.~\ref{tab:obs}), although
whether it is an intrinsic effect of the position angle (i.e.\
uncorrected distortion, perhaps related to charge transfer efficiency;
\citealt*{kpgp07}) or something to do with the changing set of
reference stars, we cannot determine. Overall, though, the deviations
are not greatly significant, and the proper motion is confirmed by our
analyses of various subsets of the data.  The corrected proper motion
is:
\begin{eqnarray} 
\mu_{\alpha}&=&-12.0\pm0.4\,\masyr\nonumber \\
\mu_\delta&=&\phn\mbox{+4.1}\pm0.4\,\masyr,
\label{eqn:pm}
\end{eqnarray}
where the uncertainty is a combination of the uncertainty on the
derived proper motion and the uncertainty in the shift to the
reference frame.
This proper motion is in a reference frame defined by the average
motion of the background stars in the field (this is the same
procedure used for the initial measurement of \citetalias{wm77}, which
is their Equation~2, but the different choices of references stars
means that the reference frames will not be exactly the same).
However, as we discuss below, this reference frame is not the
appropriate one for considering the degree of alignment between the
pulsar motion and the projected symmetry axis of the surrounding
nebula.

\subsection{Absolute Position}
\label{sec:absast}
We can get a position for the Crab pulsar tied to the International
Celestial Reference System (ICRS; useful for analysis of timing data,
for example) directly from 2MASS, which is tied to the
ICRS\footnote{See
\url{http://spider.ipac.caltech.edu/staff/hlm/2mass/overv/overv.html}.}
to better than $0\farcs1$.  The pulsar is listed as source
2MASS~J05343194+2200521, at position (J2000):
\begin{eqnarray}
\alpha&=&05^{\rm h}34^{\rm m}31\fs94 \nonumber \\
\delta&=&+22\degr00\arcmin52\farcs1, 
\end{eqnarray}
with quoted uncertainties of $\pm0\farcs06$ on each coordinate.  This
position has been precessed to equinox J2000, but it was actually
measured on 1997~October~18.  The 2.2 years between those dates
implies a shift due to the measured proper motion of 27$\,$mas, which is
smaller than the measurement uncertainty.  To verify the tie of the
2MASS data to the ICRS, we compared the position of 32 stars located
within $5\arcmin$ of the pulsar that were detected in both 2MASS and
the Second U.~S.\ Naval Observatory CCD Astrograph Catalog (UCAC2;
\citealt{zuz+04}), which is tied directly to the ICRS.  We find a
small residual shift ($\approx 32\,$mas) between 2MASS and UCAC2 (the
UCAC2 proper motions for these stars are not statistically
significant, so we did not include them), but this is again smaller
than the measurement error reported by 2MASS.  The 2MASS position is
consistent with the radio imaging position \citep{ht99} listed in
SIMBAD from the NRAO VLA Sky Survey (NVSS; \citealt{ccg+98}), but
should be more accurate.  A caveat, though, is that 2MASS could not
measure the pulsar separately from the optical/infrared knot located
$0\farcs7$ away \citep{hss+95}.  However, the pulsar is $>40$ times
brighter than the knot (at least at 5500$\,$\AA), and so the 2MASS
centroid should essentially be at the pulsar's position.

\subsection{Proper Motion Reference Frame}
\label{sec:ref}
Equation~\ref{eqn:pm} gives the proper motion of the Crab pulsar
relative to an ensemble of stars.  Normally, we would correct our
measured proper motion for the peculiar velocity of the Sun relative
to the local standard of rest (LSR; we take the Solar motion to be
$[U,V,W]_\odot=[10.00,5.25,7.17]{\rm~km\,s^{-1}}$, where the
uncertainties on those components are negligible compared to the other
uncertainties discussed below; \citealt{db98b}) and for the effects of
differential Galactic rotation (DGR; we use the Galactic potential
model of \citealt{kg89} to determine the rotation curve, and then
examine deviations from the rotation curve determined by
\citealt{bb93}), so that the proper motion reflects a velocity
relative to the object's local standard of rest.  However, we should
not apply these corrections blindly for the Crab pulsar.  The reason
is that the astrometric reference stars that we are using are at
distances of 1--4$\,$kpc (and possibly a bit further), comparable with
the nominal 2$\,$kpc distance of the Crab pulsar (\S~\ref{sec:cmd}).  To
first order, then, we do not need to correct for DGR or LSR motion.

To see how valid this is, we refine our distance estimate by following
\citetalias{wm77}.  They found a mean statistical parallax of
0.5$\,$mas for the reference stars from kinematic constraints, roughly
consistent with our estimate of 1--4$\,$kpc distances for the WIRC stars.
Indeed, the stars used by \citetalias{wm77} were spread over several
arcminutes and therefore are more comparable to the near-IR sample
than the \hst\ sample.  \citetalias{wm77} formed their estimate by
examining the dispersion of the reference stars about their mean
proper motion and then relating that to the expected velocity
dispersion.  We find a proper motion dispersion of
$\sigma_{\mu}=1.8\,\masyr$.  Comparing this to the expected
1-D velocity dispersion of $32\,\kms$ \citep[appropriate
for late-type main-sequence stars;][p.\ 632]{bm98}, we find a mean
distance of 3--4$\,$kpc, consistent with our photometric estimates and
with the likelihood that some of the \hst\ stars are fainter and more
distant than the near-IR stars.

Of course, the stars are not all at a single distance, but are
distributed over a range, and their contribution to the ensemble's
proper motion depends on their brightnesses and on how many
observations we have for them.  The range of distances will influence
the ensemble's mean proper motion: going from 1$\,$kpc to 4$\,$kpc increases
the correction due to DGR by $0.2\,\masyr$, while the
correction for the LSR decreases in magnitude by $1\,\masyr$.  Since the Crab pulsar is toward the anti-center the LSR
correction is more significant, and it depends inversely on distance.
We take the pulsar to be at 2$\,$kpc, which gives corrections of
\begin{eqnarray}
(\Delta \mu_\ell, \Delta \mu_b)_{\rm DGR}(2\,\mbox{kpc})&=&(+0.68,+0.04)\,\masyr \nonumber \\
(\Delta\mu_\ell, \Delta \mu_b)_{\rm LSR}(2\,\mbox{kpc})&=&(+0.47,-0.64)\,\masyr.
\end{eqnarray}
For the average stellar frame at 4$\,$kpc, the corrections
are:
\begin{eqnarray}
(\Delta \mu_\ell, \Delta \mu_b)_{\rm DGR}(4\,\mbox{kpc})&=&(+0.78,+0.03)\,\masyr \nonumber \\
(\Delta \mu_\ell, \Delta\mu_b)_{\rm LSR}(4\,\mbox{kpc})&=&(+0.23,-0.32)\,\masyr,
\end{eqnarray}
 so the net corrections are 
\begin{eqnarray}
(\Delta \mu_\ell, \Delta \mu_b)_{\rm
DGR}(4\mbox{\,kpc}\rightarrow 2\,\mbox{kpc})&=&(-0.10,+0.01)\,\masyr \nonumber \\
(\Delta \mu_\ell, \Delta \mu_b)_{\rm LSR}(4\,\mbox{kpc}\rightarrow
  2\,\mbox{kpc})&=&(+0.24,-0.32)\,\masyr. \nonumber \\
 & &
\end{eqnarray}
 In equatorial coordinates these corrections are 
\begin{eqnarray}
(\Delta \mu_\alpha,
\Delta \mu_\delta)_{\rm DGR}(4\mbox{\,kpc}\rightarrow 2\mbox{\,kpc})&=&(-0.05,+0.09)\,\masyr \nonumber \\
(\Delta \mu_\alpha, \Delta
\mu_\delta)_{\rm LSR}(4\mbox{\,kpc}\rightarrow
2\mbox{\,kpc})&=&(-0.14,-0.37)\,\masyr. \nonumber \\
 & &
\label{eqn:corr}
\end{eqnarray}
These corrections must be subtracted from the proper motion we found
in \S~\ref{sec:pm}, so the proper motion of the pulsar relative to its
local standard of rest is $\mu_\alpha=-11.8\,\masyr$,
$\mu_\delta=+4.4\,\masyr$.  Of course, these distances are uncertain.
Assuming a 0.5$\,$kpc uncertainty in the distance to the pulsar, and
considering the range of 3$\,$kpc to 5$\,$kpc for the reference stars,
we found that the correction overall varied by $\pm0.1\,\masyr$ from
the nominal values in Equation~\ref{eqn:corr}, so compared to the
other sources of uncertainty this is a minor effect.  We note that the
uncertainties here exceed formal 1-$\sigma$ confidence intervals,
since the distance intervals represent the full range of plausible
distances.  We also note that we have chosen a particular formulation
of the Galactic rotation curve through the potential of \citet{kg89}:
other formulations (flat rotation curves, use of Oort constants, etc.)
give slightly different results for the DGR corrections.  In
particular, without a velocity that varies as a function of distance,
the correction from $4\,$kpc to $2\,$kpc is identically 0.  Using
other choices (e.g., the rotation curve of \citealt{bb93}) changes the
correction by a small amount on an absolute scale, typically
$\pm0.1\,\masyr$ (other determinations of the LSR corrections agree to
within 10\%).  We include this as an additional uncertainty.

However, more significant uncertainties come from our assumption that
the rotation curve toward the outer Galaxy is both well-known and
well-behaved, when it is neither.  Overall, the rotation curve of the
outer Galaxy has line-of-sight random variations of $\pm 5\,\kms$ \citep[][and references therein]{bb93}, which accounts for
the deviations of individual locations from the bulk velocity.  The
curve itself is poorly measured toward the anticenter, but this value
should be relatively independent of position.  So we would expect 1-D
variations of $\pm 5\,\kms$ on top of the circular
velocity, implying a proper motion uncertainty of $0.5\,\masyr$ at 2$\,$kpc.  The proper motion is
\begin{eqnarray}
\mu_\alpha & = & -11.8 \pm 0.4 \pm 0.5\,\masyr \nonumber \\
\mu_\delta & = & \phn\mbox{+4.4} \pm 0.4 \pm 0.5\,\masyr,
\end{eqnarray}
compared to the local standard of rest of the Crab pulsar, where the
first uncertainty is the measurement uncertainty (Eqn.~\ref{eqn:pm})
and the second is from the reference frame uncertainties.

Now, if our goal is to compare the proper motion vector to the
projected orientation of the torus axis, we need to account for yet
another zero point uncertainty, namely the unknown velocity of the
Crab's progenitor.  The progenitor was presumably not stationary with
respect to the local standard of rest at that position, and any
peculiar velocity should remain after the explosion.  Therefore a
final zero-point uncertainty comes from the unknown peculiar velocity
of the Crab pulsar's progenitor.  First, there are bulk streaming
motions in the outer Galaxy with line-of-sight magnitude $\approx
12\,\kms$ \citep{bb93}.  Beyond this, early-type stars have 1-D
velocity dispersions of $\sim 10\,\kms$ \citep{bm98}.  Together these
give into a 1-D velocity uncertainty of $16\,\kms$, which,
combined with the $0.5\,\masyr$ uncertainty discussed above,
translates into an uncertainty of $1.7\,\masyr$ at 2$\,$kpc, with a
range of 1.4--2.3$\,\masyr$ for a distance range of 1.5--2.5$\,$kpc
(i.e.\ an uncertainty on the uncertainty).  We assign a conservative
value of $2.0\,\masyr$ (considering the uncertainty on the distance as
well as the reference frame velocities) for the systematic uncertainty
due to combination of the reference frame effects, but also note that
the progenitor's velocity could have been much larger (see below).  We
then find a proper motion in the reference frame of the progenitor
star of:
\begin{eqnarray}
\mu_\alpha & = & -11.8 \pm 0.4 \pm 2.0\,\masyr \nonumber \\
\mu_\delta & = &\phn\mbox{+4.4} \pm 0.4 \pm 2.0\,\masyr,
\end{eqnarray}
where again the two uncertainties are from the measurement and the unknown
reference frame, as discussed above.  This proper motion has a
magnitude of $\mu=12.5\pm0.4\pm2.0\,\masyr$ at an angle of
$290\degr\pm2\degr\pm9\degr$ (east of north), or a transverse velocity
of $120\,\kms$ for a distance of 2$\,$kpc; the velocity is
quite uncertain, both due to the uncertain frame of the proper motion
and the distance uncertainty.

\section{Discussion and Conclusions}
\label{sec:conc}

While we assumed a velocity of $\pm10\,\kms$ for the Crab
pulsar's progenitor, it is possible that the progenitor itself had a
much larger space velocity.  This was proposed early on by
\citet{mink70}, who considered the high proper motion of the pulsar to
be a relic of the progenitor's velocity and that the progenitor itself
was a runaway star (e.g., \citealt{b61}) from the Gem~OB1 association.
However, \citet{mink70} later dismissed this hypothesis since he did
not believe that the supernova was a type~II explosion.  \citet{ggo70}
have a similar idea, where they consider the Crab pulsar and the
nearby pulsar B0525+21 (two of the first pulsars to be discovered;
\citealt{sr68}) as former binary companions ejected from the Gem~OB1
association; \citet{hla93} attempted to measure the proper motion of
PSR~B0525+21 but did not find a statistically significant result.
Later analyses, such as \citet{pols94} and \citet{md01}, have revived
the hypothesis that the progenitor had a large ($>100\,\kms$) space velocity, with \citet{pols94} arguing that the large
height below the Galactic plane (200$\,$pc for a distance of 2$\,$kpc, which
is larger than the scale-height of OB stars; \citealt{reed00};
\citealt*{ecca06}) and the presumed evolutionary state of the
progenitor (inferred from the elemental and velocity structure of the
Crab Nebula) suggest that the Crab pulsar was formed from the second
explosion in a binary system, and that it had a large space velocity
from the first explosion.  Currently we cannot say definitively
whether or not the progenitor had a significant space velocity and
must therefore treat this as an overall uncertainty on our whole
analysis.

\subsection{Location of the Explosion Center}
A number of authors have estimated
the ``divergent point'' for the Crab Nebula: the point from which all
of the filaments seem to traveling outwards, which is presumed to be
the center of the explosion.  Among the more reliable measurements are
those of \citet{trimble68}, \citetalias{wm77}, and \citet[][who also
review the situation]{nugent98}, which we list in Table~\ref{tab:exp}.
Most of these determinations trace the filaments back and find
best-fit dates for the explosion of $\approx 1130$~CE instead of the
commonly accepted 1054~CE \citep{sg02}, with the difference caused by
unmodeled acceleration of the filaments.

\begin{deluxetable*}{l c c l c c}
\tablecaption{Comparison Between Our Proper Motion and Divergent Points\label{tab:exp}}
\tablewidth{0pt}
\tabletypesize{\footnotesize}
\tablehead{
\colhead{Reference} & \mc{2}{c}{Divergent Point (J2000)\tablenotemark{a}} &
\colhead{Divg.\ Date\tablenotemark{b}} & \colhead{$\Delta
  r_{\rm min}$\tablenotemark{c}} & \colhead{$\Delta r$(1054 CE)\tablenotemark{d}}\\ \cline{2-3}
 & \colhead{$\alpha$} & \colhead{$\delta$} & \colhead{(CE)} &
\colhead{(arcsec)} &\colhead{(arcsec)} \\
}
\startdata
\citet{trimble68} & $05^{\rm h}34^{\rm m}32\fs72\pm0\fs12$ &
$+22\degr00\arcmin47\farcs5\pm1\farcs4$ & $1067\pm138$ & $0.5$ & 0.6\\
\citetalias{wm77} & $05^{\rm h}34^{\rm m}32\fs67\pm0\fs06$ &
$+22\degr00\arcmin47\farcs6\pm0\farcs9$ & $1114\pm78$ & 0.7 & 1.0\\
\citet{nugent98} & $05^{\rm h}34^{\rm m}32\fs84\pm0\fs12$ &
$+22\degr00\arcmin48\farcs0\pm1\farcs3$ & \phn$947\pm138$ & 0.6 & 1.5\\
\tableline
This work\tablenotemark{e}  &$05^{\rm h}34^{\rm m}32\fs74\pm0\fs03$ &
$+22\degr00\arcmin47\farcs9\pm0\farcs4$ & 1054 & \nodata & \nodata\\
\enddata
\tablenotetext{a}{The positions of the divergent point according to
  \citet{trimble68} and \citet{nugent98} were 
  computed from the offsets given in 
  \citet{nugent98} between the divergent point/explosion center and
  the star $5\arcsec$ to the north-east of the pulsar, whose position
  we take to be: $\alpha=05^{\rm h}34^{\rm m}32\fs17$,
  $\delta=+22\degr00\arcmin56\farcs0$ from 2MASS (the star is
  2MASS~J05343217+2200560).  For 
  \citetalias{wm77}, we take the
divergent point directly from their paper (Eqn.~17).}
\tablenotetext{b}{Date of closest approach between our proper motion
  vector projected backwards and the divergent point.  The
  uncertainties are only measurement uncertainties --- no reference
  frame uncertainties are included.}
\tablenotetext{c}{Closest approach between our proper motion projected
  backward and the estimated divergent point of the filaments.}
\tablenotetext{d}{Distance between our proper motion projected
  backward and the divergent point for 1054~CE.}
\tablenotetext{e}{Not truly a divergent point, but rather the
  location of the pulsar projected back to 1054~CE.}
\end{deluxetable*}

With the proper motion that we derive, the explosion centers from the
literature, and the nominal explosion date of 1054~CE, we can perform
three tests: we have a time, a displacement, and a velocity, and we
can use any two of those to estimate the third.  First, we can measure
how close our proper motion comes to the various explosion centers for
the nominal explosion date, and we give these values in the last column
(``$\Delta r$(1054~CE)'') of Table~\ref{tab:exp}.  Second, we can
compute the dates of closest approach between our projected proper
motion vectors and the explosion centers, which serve as our own
estimates of the explosion dates. The uncertainties on those values
are dominated by the uncertainties of the divergent point measurements
($\sim 1\arcsec$), and the values are given in the ``Divg.~Date''
column of Table~\ref{tab:exp}, with the approach distances for those
dates given in the ``$\Delta r_{\rm min}$'' column.  Finally, we can
compute our own estimate for the explosion position, taking our proper
motion and assuming the date of 1054~CE, and we give this in the last
row of Table~\ref{tab:exp}.

In general, all of the values --- the approach distances for 1054~CE,
the explosion dates, and our inferred explosion center --- are
consistent at better than 1-$\sigma$.  The first two elements are
largely consistency checks: this shows that our proper motion is
consistent with the independent divergent point estimates, and that
our reference frame corrections are consistent (although they were not
explicitly the same).  As discussed in \citetalias{wm77}, the location
of the divergent point depends on the choice of reference frame, so we
cannot address any of the larger reference frame uncertainties.  We
note, though, that in contrast to our proper motion that approaches
the divergent points with distances of $<1\arcsec$, the proper motion
of \citetalias{nr06} does approximately a factor of 3 worse.  The
final element that we have computed, our own estimate for the
explosion center, is more precise than previous estimates by a factor
of $\sim$2--3 in each axis.  This may serve to help constrain future
measurements of the filament motions and acceleration, as its
independence from the filaments themselves should improve the
reliability of the measurements.

\subsection{Spin-Kick MisAlignment}
\citet{nr04} fit a model to the torus seen in the \hst\ data, and find
a best-fit torus symmetry axis of $304.0\degr\pm0.1\degr$, which is
the projection of the spin axis on the plane of the sky.  This implies
a projected misalignment of $14\degr\pm2\degr\pm9\degr$, as seen
Figure~\ref{fig:wfc}.  This projected misalignment is less than that
found in \citetalias{nr06}, and is significant if one only considers
the measurement uncertainty: with all of the uncertainties, the
misalignment is consistent with a broad range of values, including zero.

Perhaps the next best case of a pulsar wind nebula giving the
projected spin axis is that of the \object[PSR B0833-45]{Vela pulsar},
where the proper motion (corrected for Galactic rotation and solar
motion) is $45\pm1.3\,\masyr$ at a position angle of $301\degr\pm
2\degr$ \citep{dlrm03}.  The symmetry axis of the torus is at a
position angle of $310.6\degr\pm0.1\degr$ \citep{nr04}, giving the
projected misalignment of $10\degr\pm2\degr$ \citep{nr07}.  For this
system, the proper motion reference frame and corrections should be
better defined than those for the Crab: the proper motion is measured
in the radio, so the reference sources are at infinite distance; and
the Vela pulsar is reasonably close (with a distance measured through
geometric parallax) and not located at the Galactic anti-center, so
the effects of Galactic rotation are much better understood.  However,
\citet{nr04} still fail to include any allowance for the unknown
velocity of the progenitor: a $\pm 10\,\kms$ velocity at a distance of
$287\,$pc is $\pm7\,\masyr$ (or an angular uncertainty of $\sim
9\degr$), so again this completely dominates the measurement
uncertainty and makes the degree of misalignment consistent with zero.

The two examples considered here, the Crab and Vela pulsars, while the
two best X-ray tori \citep{nr04}, may not be good cases for computing
misalignment.  This is because they are both moving at smaller
velocities than the average pulsar population ($120\,\kms$ and
$61\,\kms$, vs.\ $\sim400\,\kms$; e.g., \citealt{hllk05,fgk06}), so the
unknown velocity of the progenitor has a correspondingly greater
contribution.  In fact, there is a  bias in favor of tori
being found around relatively slow pulsars.  This is because the
transition from a ``bubble'' pulsar wind nebula (PWN) with a
torus to a bow-shock PWN \citep[see][]{gs06} occurs when the pulsar has
traveled roughly 68\% of the distance from the center of the of the
supernova remnant to its edge \citep*{vdsdk04}, largely independent of
the pulsar's velocity.  So slower pulsars will spend longer in the
bubble/torus phase.  For faster moving pulsars, whose proper
motions and projected spin axes are not as well determined (in general
they are further away), the uncertainty will be closer to $1\degr$ and
will not dominate over the measurement uncertainties.  In addition, if
the model of \citet{nr07} is correct, we would expect the faster
moving pulsars to be intrinsically closer to alignment.  We also note that, for
pulsars moving at $>200\,\kms$, all of the reference frame
uncertainties discussed here will lead to uncertainties of $<10$\% on
the space velocity: this will typically be less than the uncertainty
on the distance.  Therefore, in studying the magnitude of pulsar
velocities (or of the pulsar population) the reference frame
uncertainties will not be significant.

However, we can also look at the situation from the other side.  If
the spin and kick axes were perfectly aligned in the reference frame
of the progenitor's motion, we could still (erroneously) infer a
misalignment because of a high progenitor space velocity.  In
practice, though, it is difficult to disentangle the effects of
intrinsic and apparent misalignments for single objects.  A further
complication comes from the fact that all alignments are examined only
in projection: inclinations of nebulae can be estimated (although not
directly measured), but radial velocities of pulsars are entirely
unknown, and without the third dimension any observed alignments could
still be coincidences.

A number of other pulsar wind nebulae also have symmetry axes
\citep[e.g.,][]{pksg01,hgh01,nr04}, although the lower fluxes and
larger distances make most of these hard to observe in detail.  There
are other situations where alignments are inferred but not measured
directly: for instance, using an offset from the center of a supernova
remnant to derive a kick direction (this approach can introduce
substantial systematic errors of its own; see \citealt{gcs+06}), and
deriving a rotation axis from fitting radio polarization data
\citep{drr99,lcc01,rn03,wlh06,rankin07}.  For those systems with
constrained proper motions and rotations axes, \citet{nr04} find
projected misalignments of $\sim 10\degr$, although most are
consistent with zero.  \citet{wlh06} and \citet{nr07} find similar
misalignments for a larger sample using polarization data (also see
\citealt{drr99}), but especially if we restrict the sample to the
younger pulsars (where Galactic acceleration should not have modified
the initial velocity) then the conclusions are similar to \citet{nr04}
(also see \citealt{jhv+05}).  This suggests that, at least
statistically, there still is a considerable degree of alignment
between projected spin axes and proper motions.  From this,
\citet{wlh07} and \citet{nr07} argue that the asymmetries experienced
by proto-neutron stars following core-collapse simultaneously can
impart these stars with both kick and spin, and that these asymmetries
consist of a stochastic ensemble of thrusts, each long enough to
result in rotational averaging of the resultant linear momentum
vector. The uncertainties discussed in this paper illustrate the
difficulties in measuring precision alignments (or lack thereof) in
any individual object. Further progress in these studies, for example,
via detailed analyses of how the degree of alignment depends on
parameters such as space velocity and surface magnetic field strength,
is best achieved by adding to the total number of pulsars with
information on the orientations of both spin and kick.


\acknowledgements We thank a referee for a thorough reading, C.-Y.~Ng
for helpful comments, and M.~Bietenholz for assistance in clarifying
some of the reference frame issues.  Support for this work was
provided by the National Aeronautics and Space Administration through
Hubble award AR-10667.01.  Partial support for DLK was also provided
by NASA through Hubble Fellowship grant \#01207.01-A awarded by the
Space Telescope Science Institute, which is operated by the
Association of Universities for Research in Astronomy, Inc., for NASA,
under contract NAS 5-26555.  B.~M.~G.\ acknowledges the support of
NASA through LTSA grant NAG5-13032.  S.~C.\ acknowledges support from
the University of Sydney Postdoctoral Fellowship Program.  This
research has made use of the SIMBAD database, operated at CDS,
Strasbourg, France.  This research has made use of SAOImage DS9,
developed by the Smithsonian Astrophysical Observatory.


{\it Facilities:} \facility{Hale (WIRC)}, \facility{HST (WFPC2, ACS)}

\appendix
\section{Prospects for Parallax}
\label{sec:par}

Given the importance of the Crab pulsar in our understanding of
neutron stars, its precise distance remains a surprisingly open
question.  A trigonometric parallax has not yet been measured for the
pulsar.  From the dispersion of its radio pulses and a model of the
Galactic electron density distribution \citep[NE2001,][]{cl02},
PSR~B0531+21 has an inferred distance of 1.7$\,$kpc (a distance range of
1.4---2.0$\,$kpc).  \citet{t73} estimated a range of distances between
1.4 and 2.7$\,$kpc based on a variety of lines of evidence, and the
nominal distance to the Crab pulsar and its nebula is quoted as $2.0
\pm 0.5\,$kpc.  Precise measurements of the times of arrival of radio
pulses have been used to measure radio pulsar parallaxes, but such
measurements require exceptional rotational stability, generally seen
only in some recycled pulsars \citep[e.g.,
PSR~J0437$-$4715;][]{vbb+01}. The young Crab pulsar has noisy timing
residuals and shows rotational glitches \citep*{wbl01}, ruling out such
an approach to astrometry.  Thus a parallax (and proper motion) for
the Crab pulsar must rely on imaging at some wavelength range, and
radio VLBI or optical observations with space telescopes are currently
the most plausible approaches.

From a purely numerical perspective, it appears that one should be
able to use \hst\ observations of the Crab pulsar, as described in
this work, to measure its parallax.  After all, for bright stars the
ePSF measurements and distortion solution are accurate to $0.01\,$pixel
in an individual exposure \citep{ak04,ak06,kvka07}, which is
$0.25\,$mas for the ACS/High Resolution Camera (HRC) and $0.5\,$mas for
the ACS/WFC.  With its distance around 2$\,$kpc, we expect a parallax
near 0.5$\,$mas, so with a sufficient number of exposures this should be
measurable in principle.  However, there are two limiting factors.
First, it seems that for the brightest stars systematic effects
prevent the combination of individual exposures from reducing the
astrometric uncertainty by the square-root of the number of exposures
as one might expect (see, e.g., \citealt{kvka07}).  Second, unlike in
radio interferometry where the parallax is measured relative to
quasars and radio galaxies at essentially ``infinite'' distance, we
must measure relative to other stars in our Galaxy that are at finite
distances.  For our measurement of the parallax of a neutron star at
$\approx 350\,$pc \citep{kvka07}, we found that most of the reference
stars were at 1--2$\,$kpc and therefore had parallaxes at the
0.5--1.0$\,$mas level.  If we had ignored them, our parallax measurement
would have been biased by the weighted mean of the parallaxes of the
reference stars, or $\approx 0.5\,$mas, which is quite significant.
Luckily, we had enough photometry of the field that we were able to
determine photometric parallaxes for the reference stars.  While not
very accurate individually, they were sufficiently close to the true
values (as we measured from our astrometry) to allow the correction of
the ensemble of stars and the removal of the parallactic bias.

From our color-magnitude diagram, we see that the background stars for
the Crab pulsar are largely at 1--4$\,$kpc, although there may be some at
larger distances.  Therefore the mean parallax of the background
sources is likely $\gsim 0.25\,$mas, or approximately one half of the
expected parallax of the Crab pulsar.  This is a very significant
correction and in order to measure the parallax of the pulsar with any
significance we would need to know this bias to better than $10$\%.
This might be possible with more detailed photometry (unfortunately,
while the field is frequently observed by many facilities, most use
narrow-band filters that are not suitable for spectral typing) or
limited spectroscopy, but it would require a dedicated set of
observations.

Overall, then, the prospects for an optical astrometric parallax do
not seem to be very good.  First, we would need a number of additional
astrometric \hst\ observations, likely with the ACS/HRC (again, this
may not be possible due to the recent failure of this instrument),
where the Crab pulsar is not saturated, but this will of course mean
that we detect fewer reference stars (due both to the limited field of
the ACS/HRC and the shallower exposures).  Since we are worried about
accuracy at the $<0.01\,$pix level, we must also attempt to refine our
estimate of the distortion solution and ePSF, which are difficult with
relatively sparse fields like this one.  We must also be confident in
all of our systematics at this level.  Second, we need a good number
of multi-band photometric observations to measure reliable photometric
parallaxes for at least the majority of the background stars; since we
are observing out of the Galaxy, we must be able to distinguish
between solar metallicity stars in the disk and low-metallicity stars
in the halo.  The upcoming Wide Field Camera 3 (WFC3) on \hst\ will have
\citet{stromgren66} $uvby\beta$ filters that greatly aid in stellar typing,
but this may not be enough. It is therefore our opinion that an
optical parallax is unlikely with current instruments, although it may
be possible with  future instruments such as the \textit{Space
Interferometry Mission} or \textit{GAIA}.

Very Long Baseline Interferometry (and specifically the Very Long
Baseline Array) has been used to measure the proper motions and
parallaxes of a number of neutron stars, including some which are both
weaker and more distant than the Crab pulsar \citep[see,
e.g.][]{bbgt02,cvb+05}.  Such observations would provide astrometry
referenced to distant extragalactic quasars, eliminating uncertainty
due to the reference frame (c.f.\ our discussion in \S2.2, \S2.5, and
above), although we would still require DGR and LSR corrections.
However, the Crab pulsar is embedded in an extremely radio-bright
nebula, which dominates the system temperature of radio telescopes and
thus limits the signal-to-noise ratio of radio interferometric
observations.  Coupled with the absence of a suitable extragalactic
reference source nearby, the large increase in system temperature has
limited attempts to measure a precise proper motion and parallax with
the VLBA.  Simply adding sensitivity (by increasing the collecting
area, bandwidth, or integration time) does not address these
limitations.  We note that the $\sim$6\arcmin\ size of the Crab nebula
is comparable to the size of the primary beam of the 25-m VLBA
antennas ($\sim$9\arcmin\ at 5$\,$GHz).  A telescope consisting of
smaller dishes (such as the current Reference Design for the
\textit{Square Kilometer Array}\footnote{See \textit{SKA} Memorandum
\#69 at \url{http://www.skatelescope.org/PDF/memos/69\_ISPO.pdf}.})
would have a wider field of view and suffer a proportionately smaller
decrease in signal-to-noise ratio for the same total collecting area
when observing a source as strong as the Crab nebula.  The wider field
of view and higher sensitivity would also provide suitable astrometric
reference sources.  Future radio telescopes with continent-sized
baselines may thus enable a VLBI parallax for the Crab pulsar.

It is unfortunate that the Crab pulsar, a subject of such intense and
detailed investigation for so long, remains just beyond our current
astrometric capabilities.  However, the next generation of optical and
radio telescopes should allow the measurement of a trigonometric
parallax to this object, finally settling questions about its distance.




\clearpage

\end{document}